# Blockchain in the management of science:

# conceptual models, promises and challenges

Blockchain has received much attention recently, due to its promises of verifiable, permanent, decentralized, and efficient data handling. In 2017-2019 blockchain (and associated technologies such as smart contracts) has progressed beyond cryptocurrencies, and has been adopted in banking, retail, healthcare, and other fields. This study critically examines recent applications of blockchain in science, touching upon different stages of research cycle – from data management to publishing, peer review, research evaluation and funding. The paper is based upon a review of blockchain projects, relevant literature, a set of interviews and focus groups with startup founders, scholars, librarians, IT experts from the EU, USA, Russia, and Belarus. Proponents of blockchain for science present this technology as a tool to make science free from bias, red tape, data fraud, as well as provide innovative means to secure financial backing for new ideas. However, these projects face a set of challenges. One issue concerns introducing crypto economy (with its financial incentives) into science, a field that emphasizes disinterested and non-pecuniary pursuit of truth. Another source of concern relates to the ongoing conflict between the principle of decentralization inherent to blockchain and the practice of forcing it from above, by the state and other centralized entities.

Highlights:

- Blockchain technologies have provided a solution to challenges on all stages of the research cycle, from data sharing to publishing and funding.
- This study examines use cases of blockchain in science, drawing upon systematic review of extant projects and a set of sociological interviews with stakeholders.
- There is a tension between principles of decentralization and disintermediation inherent to blockchain, and the practice of implementing DLT solutions in a centralized way, by states and universities.
- Introduction of token economy and crypto incentives into academia, while providing scientists with new funding opportunities, runs the risk of corrupting the operation of science as a collective disinterested search for new knowledge.

Artyom Kosmarski (Артём Космарский), Laboratory for the Study of Blockchain in Education and Science (LIBON), State Academic University for the Humanities (GAUGN), Moscow, Russia.

kosmarski@gaugn.ru

**Funding**

This work was supported by the Russian Foundation for Basic Research (RFBR), grant number 18-29-16184.

Artyom Kosmarski has done extensive research in the field of STS and anthropology of science. Blockchain caught his attention as a hotbed of various and sometimes unpredictable solutions to the grievances of academic community, as well as a pathway to a more flexible, grass-root and reputation-based governance of science. Right now Artyom is studying the implementation of smart contracts in universities (research project funded by the Russian Foundation for Basic Research) and investigating the potential of token-curated registries as a basis for reputation metrics. He is also deputy head of the Laboratory for the Study of Blockchain in Education and Science (LIBON) at the State Academic University for the Humanities (GAUGN), Moscow.

## 1. Introduction[1]

The rapid development of blockchain (distributed ledger technology, DLT), building upon the success of Bitcoin, has made it a promising and potentially disrupting technology in a multitude of fields. In 2017-2018 blockchain (and associated technologies such as smart contracts) has been adopted in banking, retail, supply chain management, healthcare, even public administration (Chung-Shan Yang 2019; Tönnissen & Teuteberg 2019; Angraal, Krumholz, & Schulz 2017). The appeal of blockchain to industry and academia builds upon its promise to make data stable, transparent, and decentralized.

A distributed ledger is a set of blocks connected by cryptographic tools in such a way as to make it impossible to change the content of one block without interfering with all other blocks. In this database, transactions are transparent, and the information about them is stored on the computers of all the participants. This decentralization prevents altering or destroying the data by infiltrating the core of the system.

However, it is not the data handling but the social appeal of blockchain (Atzori, 2015) that has attracted the attention of academia. Principal advantages of blockchain in this perspective, apart from stability and verifiability of data, is the guarantee of trust in the trustless environment and successful peer-to-peer interactions without the need for a central governing body ("the third party"). These features dovetail with the logic of modern science: it is international, decentralized – there is no governing body that decides everything (Polanyi, 1962) – and develops thanks to networks of trust within the academic community (peer review system and invisible colleges (Crane, 1972). The analogy was not lost on a few early observers: "Scientific information in its essence is a large, dynamic body of information and data that is collaboratively created, altered, used and shared, which lends itself perfectly to the blockchain technology" (van Rossum, 2017: 8).

---
[1] This preprint was submitted to the "Technology in Society" journal in October 2019.

Infrastructure and management of science, on the contrary, is fraught with bias, red tape, data manipulation, and non-transparent black boxes (Bunge, 1963) on every level, from research data handling to funding, publishing, and research evaluation. How does the selection of reviewers for a manuscript submitted to the journal go? Who assesses the quality of a scientist's work? How and by whom are the recipients of a grant determined? The opacity of these processes, their inertia, bureaucratization, corruption often cause discontent of scientists (Rockwell, 2009). Scientific research itself is also full of black boxes: p-hacking, ex post facto hypothesizing, and outright scientific fraud (Head, Holman, Lanfear, Kahn & Jennions, 2015; Fanelli, 2009). Thus, introducing blockchain into science might at least crack open some black boxes and make the processes inside them more transparent, reliable, and efficient.

However, as science is not strictly business and still depends on public funding, blockchain and crypto projects were slow to emerge in this field. The first tentative projects and academic publications appeared in 2015-2016. Then, at the peak of the ICO boom in 2017, several startups promised to solve all the problems of science with their tokenized economy, some of them ending up as scam (globex.sci, scientificcoin.com), others just petering out for the lack of funds (scienceroot.com). Eventually, blockchain for science proved more workable as a set of solutions for specific processes (peer review, data storage, funding, etc). At the same time, there has emerged an extensive network of scholars, IT-experts and crypto-enthusiasts, many of whom have joined Blockchain for Science association (blockchainforscience.com) and research groups in Europe and the USA. Finally, corporations began to enter this field – IBM, for instance, has obtained a patent for a platform for the collection and analysis of scientific data on the blockchain (Suberg, 2018). Therefore, early experience of blockchain adoption in science management and organization makes it possible to review the challenges and barriers to its implementation.

My paper is based upon a critical review of blockchain projects, relevant literature, a set of interviews (N=22) and focus groups (N=3) with startup founders, scholars, librarians, IT experts, and blockchain evangelists from EU, the USA, Russia, and Belarus. The interviews were taken in November 2018 – September 2019 in the framework of an interdisciplinary research project "Smart Contracts as an Instrument for Regulation and Administration of Science", supported by the Russian Foundation for Basic Research (RFBR). Participants were recruited through snowball sampling in social networks (the objective was to reach a varied sample of experts that have had an experience of organizing or evaluating DLT projects for science). The interviews were conducted in person or by Skype, and lasted on average 60 minutes. Focus groups (group discussions of a set of particular topics) were conducted in three universities, with professors, staff, and post-graduate students who have demonstrated awareness and interest in the issues of blockchain. The focus groups lasted 90 minutes.

The paper is organized as follows: the next two sections present an overview of blockchain-based solutions for research data verification, academic publishing, and peer review. Section four outlines the challenges of introducing crypto economy (with its financial incentives and profit-seeking behavior) into science, a sphere that emphasizes disinterested and non-pecuniary pursuit of truth. Section five discusses the ongoing conflict between the principle of decentralization inherent to blockchain and the practice of forcing blockchain systems from above, by the state and other centralized entities. Finally, I discuss the strengths and weaknesses of blockchain solutions in the context of digital economy and b2c IT products.

## 2. Blockchain and research data

The replication crisis is arguably the most wide-scale crisis afflicting contemporary academia, particularly the social and life sciences. In many disciplines, the impossibility of reproducing experiments has brought about retractions of seminal studies. In psychology, for instance, researchers failed to reproduce the results of 59 of the 98 well-known works (Baker, 2015). A survey conducted by Nature in 2016 showed that more than 70% of the 1576 scientists surveyed tried and failed to replicate their colleagues' experiments (Baker, 2016). In the biomedical sciences, the problem is no less acute (Ioannidis, 2005).

However, non-reproducibility is only the tip of the iceberg. Science suffers from errors and distortions at all stages of the research cycle, from dubious data collection and protocol procedures to the distorting of evidence to fit the hypothesis (Simmons, Nelson, Simonsohn, 2011), as well as p-hacking (Head, Holman, Lanfear, Kahn & Jennions, 2015). The pressure to publish and journals' preference for positive rather than negative results (positive-results bias, see (Sackett, 1979) have lead to the formation of the so-called false chain of research: new studies are based on untested (and probably erroneous) old ones. However, these problems are not caused exclusively by the malicious intent of fraudulent scientists. There are many reasons for false data, from inevitable inaccuracies in large collaborations to salami-slicing (need to publish even small results as soon as possible in order to extend funding (Fochler & Sigl, 2018).

Blockchain solutions could fix that by making the research cycle open and transparent, and by facilitating data sharing. Discoveries may be rapidly recorded in the distributed registry, indicating the authorship and date of discovery. Time-stamping on the blockchain acts as a novel way to protect ideas without resorting to (slower) patents and publications (Benchoufi & Ravaud, 2017; Moehrke, 2016). Furthermore, blockchain allows for tracking the entire scientific project from hypothesis to data collection and further analysis. The stability of data on the blockchain is essential here (all changes are trackable). By uploading the data into such database and making it open to a broader academic community, researchers will no longer be able to tamper with it, rigging the data to achieve necessary results, remove outliers, etc. (Blockchain for Open Science and Knowledge Creation: 13-18)

However, regardless of the fact that virtually every text on blockchain in science praises the potential of this technology to overcome the replication crisis and to foster open, fraud-free science, little progress has been made in this direction. The first reason is technical: recording vast amounts of research data on the blockchain requires considerable computation power, and for most blockchains currently in use it would be a long and expensive process. Second, transparency and availability of research data on DLT require mass support of scientists. If only a handful of enthusiasts pursues this task, it will not become the norm for the academic community, and this would compromise the whole project.

Next, making all research data transparent and verifiable takes researchers' time without bringing apparent benefits in their careers. Therefore, open science on the blockchain could probably be implemented only from above, forced upon the scientists by public agencies and foundations. This fact is grudgingly acknowledged even by blockchain enthusiasts: "*The state should fund Blockchain and related digital infrastructure, there is no other option. It is a difficult and complex task. You can't do these things through short-term grants or private money*" (D., decentralized data startup founder). The big public agencies in Europe and other Western countries are still too distrustful of DLT (at this early stage of their development). Also, individual researchers lack resources and initiative to "blockchainify" their workflow. Therefore, this trend of DLT for research data is rather slow to develop.

Another avenue for open science on the blockchain lies in encouraging transparency through material incentives. In other words, a DLT platform with its tokens and rules that make adherence to open science (e.g., open and valid data, negative results made public) lucrative for scientists. The EUREKA platform (eurekatoken.io) has advanced in this direction, promising to reward scholars with tokens for publishing both positive, negative, and uncertain results. Also, a separate token fund is allocated to pay for replication studies and experiments aimed at increasing reproducibility. The success of this initiative, however, depends on the price of EUREKA tokens on the market (after a forthcoming ICO). Even still, the pegging of open science principles to the vagaries of market speculation (the forces beyond scientists' control) is risky at best.

**3. Blockchain and academic publishing**

Academic publishing is passing through turbulent times – the crisis of subscription model, the rise of open access journals, and potentially disrupting Plan S with its forceful drive of compulsory transition to open access journals (Else, 2018). Regardless of the payment model, the publishing cycle is extremely slow: writing an article, submitting it to a journal, searching for reviewers, getting feedback, and finalizing takes months if not years (Smith, 2006). Researchers are involved in the race for priority, but the current system significantly slows down the exchange of results, let alone ideas. Finally, the process itself is deeply flawed. Double-blind review frequently turns out to be pseudo-anonymous, with little limits to reviewers' bias (Wang, Kong, et al., 2016). Also, reviewers are increasingly overworked and underpaid (Kovanis, Porcher, Ravaud & Trinquart, 2016). The need for incentives in peer review, e.g., "academic dollars" was voiced well before the advent of cryptocurrencies (Pruefer & Zetland, 2009).

The advantages of blockchain technology for solving these problems are quite obvious. Minimally, it provides notarization. Recording a text or even a draft idea on the blockchain (time-stamping) allows a scientist to assert priority and intellectual property rights, and then he or she might freely share it as a preprint. Further, the decentralization and disintermediation principles behind blockchain suggest an independent publishing platform where authors and reviewers interact directly with each other in a p2p network, with no need for excessive publishing and subscription costs. Not surprisingly, this idea has been so appealing that virtually each blockchain startup in science promised an open access platform (scienceroot.com, eurekatoken.io, pluto.network, orvium.io).

However, the success of these platforms has been so far limited, with less than a hundred papers in each: scientists prefer to publish in established journals relevant to their research communities. This trend is well-known in the sociology of innovation (Dahlin, 2014): the revolutionary benefits of new technology are insufficient to draw users away from customary practices. Moreover, running an academic journal on an automated system of smart contracts is an arduous task, and developing an efficient ecosystem for several journals with their own rules and principles is a process beyond the scope of current blockchain projects (Janowicz, Regalis, et al., 2018).

There is another way to integrate blockchain into academic publishing – by putting in into the service of corporations. "Blockchain for Peer Review", a project implemented jointly by the developer Katalysis (katalysis.io) and the Digital Science (a technology company), has followed this path. Its main goal is to develop a protocol that would allow collecting information about reviewers from publishers, storing it on the blockchain, and then making it possible to evaluate the work of reviewers while maintaining their anonymity. In other words, the distributed register allows recording the connection between the reviewer and the manuscript without revealing the author's public identity. Players such as Springer Nature, Taylor & Francis, Cambridge University Press, and ORCID are already involved in the project. However, this approach to DLT is entirely devoid of critical disrupting qualities of blockchain – decentralization, elimination of the third party (such as publishers). "Blockchain for Peer Review" has met criticism from the community, precisely because it's a narrow, technical and corporate solution that does not bring about any disruption to the broader scientific ecosystem.

## 4. Blockchain, research funding, and incentivization

Research funding is fraught with bias, cumbersome, non-transparent, and ineffective procedures – the "black boxes" mentioned above. Furthermore, scientists have to spend a great deal of their time writing reports, grant applications, and doing paperwork (Link, Swann & Bozemann, 2008). Finally, a pressing problem is the shrinking of funding. Governments are gradually moving away from large-scale research funding, hoping that business, industry, and private foundations will replace it (Mervis, 2017).

How could DLT fix these issues? First of all, an automated system of disbursement of funds with transactions on smart contracts will significantly reduce overhead costs and ease the burden on accountants, auditors, and scientists themselves. It would save them from filling out a lot of papers and make the whole process of allocation and distribution of funds more efficient. Also, a funder might set a combination of conditions (e.g., citations, articles, datasets) and peg the grant money to the fulfillment of these conditions (through smart contracts) – this approach is implemented in the DEIP blockchain ecosystem (deip.world).

However, the innovative drive of DLT lies in the fact that science funding and incentive structures may be easily changed with blockchain. "*We can experiment with new money distribution schemes, grant schemes, and that would bring about cultural change. With blockchain, things would change much quicker*" (S., blockchain evangelist). Entering the cryptocurrency sphere gives the scientist a chance to find money from investors whose interests and outlook are very different from universities. In such cases, blockchain will provide investors with a guarantee against scams and roguish projects: all the initial data and development of the research can be traced, and the allocation of funds can be pegged to the achievement of certain milestones.

Moreover, one could receive tokens if their research results are validated independently by others, or used in their future work, as an economic equivalent of citation. Thus token economy creates a potentially powerful channel for financing and implementing breakthrough ideas, even in basic science (Blockchain for Open Science and Knowledge Creation: 25)

In this sense, crypto-economic tools give more independence to scholars, opening an alternative channel of recognition and funding, giving scientists a clear economic interest to engage in the crypto economy. In the end, science would get more independent economic agents, apart from the state and big funding agencies and philanthropies. "*There we have more opportunities for more independent players, and for more intermediaries – and that is an interesting contradiction, as they say that blockchain is all about disintermediation. And we will probably have more players here, with less bureaucracy and more efficient, blockchain-based ways to redirect money in different directions*" (L., librarian).

However, these innovations are not easily adopted. Most scientists are not entrepreneurs working for themselves and pursue their goals within complex institutional structures, while the latter are not always friendly towards new technologies."*Smart contracts might efficiently indicate expenses of the lab – test animals, chemical agents, etc. – and allocate necessary funds. But what if you have an accountants office working with these expenditures, for many years, 50 people in all, who would think of firing them?*". (E., researcher). Furthermore, the dubious reputation of crypto-economic tools (ICO scams, hacker attacks targeting crypto wallets), sluggish transaction speeds of many blockchains makes university authorities doubt the practicality of introducing DLT into their finances.

However, these are merely technical issues, easily to be overcome in the near future, as the technology matures. There is a more fundamental challenge to merging the crypto-economy with science. At the heart of Bitcoin, the most successful blockchain project, as envisaged by its architect Satoshi Nakamoto, lies an incentive system. Bitcoin arranges incentives and rewards for all members of its ecosystem (miners, users, developers) in such a way that the output is stable, secure, and at the same time decentralized digital currency. It is through incentives that Satoshi Nakamoto has ensured that behavior beneficial to the common good of the system is encouraged and that harmful behavior is blocked (Nakamoto, 2008).

Incentive design principles were conceptualized as one of the key benefits of blockchain. They entered other cryptocurrencies and then got integrated into more complex and diverse projects. The developers laid down what behavior will be encouraged by the participants. The mechanism of encouragement itself is material: through tokens, which ultimately turn into money. Finally, the third important feature of incentive design is voting and decision-making by a simple majority of votes. For example, Gnosis, Augur, and other predictive markets on blockchain encourage users to make accurate predictions and bet on the results of these predictions. Steemit social network rewards popular posting with tokens (popularity is determined by the number of votes cast, i.e., upvoting). Various token-curated registries (Goldin, 2017) encourage the creation of authoritative lists (cafes, universities, media), encouraging responsible voting for or against the inclusion of a unit in the list. Thus, blockchain has made it possible to create powerful incentive machines (McConaghy, 2018).

However, incentive design might go rogue as one moves the field where living creatures, not machines, make decisions. In Bitcoin, incentives are automatic and embedded within the system so that people are not required to make decisions all the time (it is enough to just mine). When this mechanism is replicated in science, for instance, it would hardly be realistic to expect rational behavior from the actors. Humans are subject to cognitive biases, herd instincts, often prefer short-term gains to long-term benefits (Verbin, 2018)

Moreover, the "engine" of most blockchain startups is the system of tokens, which motivate scientists, vote, connect individual system nodes, and, ultimately, attract investors. The "tokenization" of science is a dubious enterprise. For scientists, the desire for recognition and the pursuit of truth – non-monetary incentives – are no less important than material ones (Jindal-Snape & Snape, 2006). Introduction of quantitative metrics and market mechanisms might corrupt science as a social institution: "*when you introduce monetary incentives into Wikipedia or peer review, you destroy them*"(L., librarian). In addition, market logic (when everyone strives to maximize their profits) atomizes the scientific community, further undermining the logic of science as the collective search for the truth, when the common goal is more important than individual career success - the Mertonian norm of disinterestedness (Merton 1973; Higginson et al. 2016). We face a dilemma: blockchain for science initiatives intend to build a self-regulating system run by scientists themselves, stimulating scientific progress in a self-governing sphere, but this new vision is based on a race for rewards and monetary incentives. The advantages of the crypto economy (new funding opportunities, more freedom) are partially offset by the reluctance of scientists to take on the role of business people. Not every researcher is ready to act as an investor or a startup manager who attracts investments for their project.

**Blockchain – democracy or coercion?**

Apart from a set of technical solutions, blockchain in science is turning into a movement within academia, with its ideology and wide-ranging goals. The impetus of this movement is mounting dissatisfaction with the oligopoly of large publishing houses, the "tyranny of metrics" (Muller, 2018) and indicators, the alarming growth of biased and non-reproducible research, the "precariatization" of scientists. Blockchain addresses these concerns with its promise to radically restructure the rules of the game in science: transparent transactions and decision-making, tokens as flexible incentives, new communities based on common rules set in code (Berg, 2017), decentralization, and an opportunity for scientists themselves to determine what is important (for example, to encourage reproducible research).

However, the **practical** applications of blockchain in science often run afoul of this ideology. Academics face vigorous time constraints and are naturally reluctant to engage in community projects that demand active participation, discussion, voting, and other obligations of participatory democracy (Pateman, 2012). "*The state is motivated to employ blockchain to investigate research fraud. The state, the university, not the scientists themselves. We planned to structure the whole process of evaluating publications and dissertations in our university on the blockchain. But the project failed because nobody was motivated. Everybody cares only about her data, and not about checking and evaluating her colleagues' input*" (A., vice-principal).

In the meantime, corporate players are beginning to take an interest in blockchain in science (Suberg, 2018; Novotny, Zhang et al., 2018). This development reflects the general dynamics of blockchain's fortunes: creators and enthusiasts of technology conceived it as a path to digital democracy, life without banks, states, and corporations, whereas DLT are becoming a tool to optimize management processes and strengthen the power of the same states and private companies. "Corporations enjoy a significant head start in the race to program their logics into mainstream blockchain applications, as well as the capacity to enact state policies that block new applications threatening future disintermediation... the corporatization of blockchain toward the ends of corporate sovereignty. (Manski & Manski, 2018).

Corporate applications aside, DLT itself, as a technology, is not so decentralized and libertarian as its creators and evangelists have portrayed it – blockchain coders and developers enjoy an advantage over ordinary users, effectually acting as sovereign authority of their platforms [Manski 156]. As one interviewee put it, "*Buterin [co-founder of Ethereum] preaches decentralization, but the meta-platform which sets the basic rules, the infrastructure on which a whole lot of independent projects and tokens is based, belongs to him. Decentralized autonomous organizations (DAOs) on Ethereum are not truly independent because they are founded on his code. The more numerous and independent these projects become, the more his monopoly is entrenched*" (O., researcher). Finally, as blockchain technologies are implemented from above, by the states eager to make use of transparency and data permanence that DLT offers, the scientists face the perspective of increased control and coercion – an extra obligation to register all their data on the blockchain, for instance. Proponents of new technology hail this as an essential step towards better science (van Rossum, 2017: 11), but from an end-user perspective, this could look more like a burden than an advantage.

**Conclusion**

Taking into consideration ambitious goals and reasonable ideas put forward by blockchain projects in science, it is slightly surprising that no platform or application has yet become an unquestionable success story. This slow tempo of innovation may be explained by legal uncertainties surrounding smart contracts and cryptocurrencies (thus making universities and public foundations wary). Another barrier – to make use of DLT projects, users should be minimally aware of the basic principles of DLT, which is rarely the case outside computer science. Last but not least, certain infrastructural conditions are to be met: fast and reliable Internet, sufficient computing power to confirm transactions, solving the problem of user identification, and access to cloud services (Rachovitsa 2018: 21). Quite a few universities, especially in developing countries, cannot ensure these conditions.

However, other trends are more favorable towards the implementation of blockchain in the management of science. In recent years, scholars all over the world have been increasingly mastering cloud applications aimed at automating different stages of the research cycle: note-taking (Evernote), collaborative writing (Authorea, GoogleDocs, Overleaf), references and citations management (Mendeley, Zotero), data exchange (Figshare, GitHub). Experimental approaches have permeated the practice of science, be it new publishing models (open access), new metrics, such as altmetrics, new forms and practices of reviewing (open peer review, collaborative peer review). Blockchain fits into this trend quite naturally. The problem is, existing DLT projects for science lack a "killer app", simple and efficient, aimed at solving one or maximum two problems evident to thousands of scientists – an approach modeling the success of the applications mentioned above. Projects under development lean toward complex solutions, ecosystems that reshape whole rules of the game. Such an ambitious approach is riskier and would succeed only if enough stakeholders throw their weight behind a specific project. "Introducing a blockchain for research and its successful adoption will depend on the collaboration between all stakeholders: funders, government, institutions, publishers, and researchers themselves, whether in their role as researcher, reviewer, editor or author" (van Rossum, 2017: 15).

One could argue that blockchain has every chance of creating new opportunities and opening new avenues in the management of science. But mere promises of a brighter digital future are not enough. To make it to the top, DLT projects in science should succeed in reaching out to individual scientists, as well as to the scores of fragmented academic "tribes" (Becher & Trowler, 2001). Successful applications should integrate seamlessly into existing practices and procedures into scientists' daily lives, should their work more comfortable rather than more demanding. Finally, they should prove to key stakeholders, especially the universities and major foundations, that blockchain provides a standard of accuracy, transparency, and reliability.

**Conflict of interest statement**

The author declares no conflict of interest.

**Funding**

This work was supported by the Russian Foundation for Basic Research (RFBR), grant number 18-29-16184.

Atzori, M. (2015). Blockchain Technology and Decentralized Governance: Is the State Still Necessary? (December 1, 2015). http://dx.doi.org/10.2139/ssrn.2709713

Baker, M. Over half of psychology studies fail reproducibility test (2015). Retrieved from Nature News. Accessed 20 October 2019. https://www.nature.com/news/over-half-of-psychology-studies-fail-reproducibility-test-1.18248

Baker, M. 1,500 scientists lift the lid on reproducibility (2016). Retrieved from Nature News. Accessed 20 October 2019. https://www.nature.com/news/1-500-scientists-lift-the-lid-on-reproducibility-1.19970

Becher, T., & Trowler, P. R. (2001). *Academic tribes and territories: intellectual enquiry and the cultures of disciplines*. (2nd ed. ed.) Buckingham: Open University Press/SRHE.

Benchoufi M, Ravaud P. (2017) Blockchain technology for improving clinical research quality. *Trials*. 18, 335. https://doi.org/10.1186/s13063-017-2035-z

Berg, C. (2017) Delegation and Unbundling in a Crypto-Democracy (July 13, 2017) http://dx.doi.org/10.2139/ssrn.3001585

Blockchain for Open Science and Knowledge Creation. Living document. Accessed 20 October 2019. https://docs.google.com/document/d/1Uhjb4K69l0bSx7UXYUStV_rjuPC7VGo0ERa-7xEsr58/edit

*Bunge, M*. (1963). A General Black Box Theory. *Philosophy of Science.* 4 (30), 346-358. https://doi.org/10.1086/287954

Crane, D. (1972) I*nvisible colleges. Diffusion of knowledge in scientific communities*. Chicago and London: The University of Chicago Press.

Dahlin, E. C. (2014). The Sociology of Innovation: Organizational, Environmental, and Relative Perspectives*. Sociology Compass,* 8, 671-687. https://doi.org/10.1111/soc4.12177

Else, H. (2018). Radical open-access plan could spell end to journal subscriptions. *Nature, 561*, 17-18. https://doi.org/10.1038/d41586-018-06178-7

Fanelli, D. (2009). How Many Scientists Fabricate and Falsify Research? A Systematic Review and Meta-Analysis of Survey Data. *PLOS One*. (May 29, 2009). https://doi.org/10.1371/journal.pone.0005738

Fochler, M., Sigl, L. (2018). Anticipatory Uncertainty: How Academic and Industry Researchers in the Life Sciences Experience and Manage the Uncertainties of the Research Process Differently. *Science as Culture*, *3 (27),* 349-374. https://doi.org/10.1080/09505431.2018.1485640

Goldin, M. (2017). Token-Curated Registries 1.0 (white paper). Accessed 20 October 2019. https://docs.google.com/document/d/1BWWC_-Kmso9b7yCI_R7ysoGFIT9D_sfjH3axQsmB6E/edit